\documentclass{article}
\usepackage{graphicx}  
\usepackage{amsmath}   
\usepackage{amssymb}   
\usepackage{bm} 
\usepackage{dcolumn}
\usepackage{color}
\usepackage{mathrsfs}
\usepackage{amsfonts}
\usepackage{varioref}
\usepackage[margin=1.37in]{geometry}
\usepackage{slashed}
\RequirePackage[colorlinks,citecolor=blue,urlcolor=magenta,linkcolor=blue]{hyperref}
\def\p{\partial}
\def\e{\epsilon}

\def\be{\begin{equation}}
\def\ee{\end{equation}}

\labelformat{section}{Section #1} 
\labelformat{subsection}{Section #1} 
\labelformat{subsubsection}{Section #1}
\labelformat{subsubsubsection}{Section #1}
\labelformat{equation}{Eq.~(#1)} 
\labelformat{figure}{Fig.~#1} 
\labelformat{subfigure}{Fig.~\thefigure#1} 
\labelformat{table}{Tab.~#1} 
\labelformat{appendix}{Appendix #1}

\title{\bf Yukawa scalar self energy at two loop and $\langle \phi^2 \rangle$ in the inflationary de Sitter spacetime}
\author{$^1$\bf Sourav Bhattacharya\footnote{sbhatta.physics@jadavpuruniversity.in} \, and \,  $^2$Moutushi Dutta Choudhury\footnote{mdchoudhury@kol.amity.edu}\\
\small{$^1$Relativity and Cosmology Research Centre, Department of Physics, Jadavpur University, Kolkata 700 032, India}\\
\small{$^2$Department of Physics, Amity Institute of Applied Sciences, Amity University - Kolkata, AA2, Newtown, Kolkata 700 135, India}}
\begin{document}
\maketitle
\begin{abstract}
\noindent
We have considered the Yukawa theory at two loop in the inflationary de Sitter spacetime, for a massless minimally coupled scalar and a massless fermion. The one loop computation for the same has been investigated in detail in the earlier literatures. The chief motivation behind this study is the fact that at one loop, the scalar self energy contains only fermions, which are conformally invariant. At two loop, there are two diagrams, each containing one internal  scalar line, thereby breaking the conformal invariance. This should result in the appearance of infrared secular logarithms in the scalar self energy.  The renormalisation of this two loop self energy  has been performed. We next compute the loop corrected coincident two point correlation function, $\langle \phi^2\rangle$, due to the self energies. The expectation value has been taken with respect to the initial Bunch-Davies vacuum. We argue that the late time secular contribution from the local or UV self energy must dominate the non-local or IR ones in the present case, from the point of view of the powers of these large logarithms of the scale factor. This corresponds to the fact that fermion lines  do not show any IR secular effect. The leading behaviour of $\langle \phi^2\rangle$ at one and two loop are respectively found to be $\ln^3 a$ and $\ln^4 a$. These are hybrids of UV and IR logarithms, where the latter originate from the massless and minimal two external scalar lines. A resummed expression for $\langle \phi^2\rangle$ has also been computed. The same is found to be bounded and decreasing monotonically with the increasing magnitude of the Yukawa coupling.  Accordingly, the dynamically generated scalar mass increases with the increasing  coupling.
\end{abstract}
\vskip .5cm

\newpage

\tableofcontents

\section{Introduction}\label{S1}
 
The  primordial cosmic inflation is a proposed phase of our very early universe characterised by a  rapid, near exponential accelerated expansion.  Such inflationary phase gives answers to the well known three   puzzles of the standard big bang cosmology, viz., the horizon problem, the flatness problem and the non-observed topological defects like the magnetic monopoles.  It also provides a suitable framework  to generate primordial quantum  density perturbations and correlation functions, as  seeds to the large scale cosmic structures we observe today in the sky. This framework predicts a nearly scale invariant power spectra consistent with observations.  We refer our reader to~\cite{Wein} and references therein for various theoretical and observational aspects of this very interesting and important phase of our early universe. 
 
In general relativity, an accelerated expansion of the spacetime requires violation of the strong energy condition. This is satisfied by some exotic matter field called the dark energy having positive energy density but negative isotropic pressure. Traditionally in the literature on inflation, this is achieved by a  scalar field, called the inflaton, slowly moving down a potential. After sufficient number of $e$-foldings, the universe  must gracefully exit the inflationary period,  specifying the current observed value of the cosmological constant. Only $10\%$ change to this value can modify the evolution of our universe greatly, known as the cosmic coincidence  problem. We refer our reader to e.g.~\cite{Tsamis, Inagaki:2005qp, Ringeval, Padmanabhan, Alberte, Appleby, Evnin:2018zeo} and references therein for various discussions  pertaining to this issue.

The de Sitter or its variant  is believed to be the metric appropriate to describe the inflationary phase. This rapidly accelerating background describes a highly non-equilibrium scenario. In such a background, understanding quantum fluctuations is an important and interesting task. Various field quantisation in de Sitter background can be seen in~\cite{Chernikov:1968zm, Bunch:1978yq, Linde:1982uu, Starobinsky:1982ee, Allen:1985ux, Allen}. In particular, the quantum fluctuations of massless but not conformally invariant  fields such as a massless minimally coupled scalar, gravitons, are of special interest in this context. Such fields do not have any de Sitter invariant Wightman functions~\cite{Allen:1985ux, Allen},   leading to appearances of logarithm of the scale factor in various quantum amplitudes. This eventually leads to the breakdown of the perturbative  expansion at late times, regarded  as the secular effect~\cite{Floratos}. The secular logarithms are chiefly related to the super-Hubble, deep infrared fluctuations at late times. They can lead to dynamical generation of the field  rest mass. For various aspects of this non-perturbative effect including mass generation, cosmic decoherence and entanglement, we refer our reader to e.g.~\cite{davis, Onemli:2002hr, Brunier:2004sb, Prokopec:2003tm, Kahya:2009sz, Miao:2012bj, Boyanovsky:2012qs, Onemli:2015pma, Prokopec:2007ak, Cabrer:2007xm, Akhmedov:2011pj, Akhmedov:2012pa, Beneke:2012kn, Akhmedov:2013xka, Boran:2017fsx, Karakaya:2017evp, Liao:2018sci, Akhmedov:2019cfd, Glavan:2019uni, Boyanovski, Bhattacharyya:2024duw, Brahma:2024ycc, Zhang:2018dvc, Zhang:2019gtg, Ye:2022tgs} and references therein.
Resummation of these secular logarithms has  been attempted in numerous ocassions, see e.g.~\cite{Burgess:2009bs, Arai:2012sh, Arai:2013jna, LopezNacir:2013alw, Burgess:2015ajz, Serreau:2013eoa, Moreau:2018lmz,  Moreau:2018ena, Youssef:2013by, Ferreira:2017ogo, Kitamoto:2018dek, Baumgart:2019clc, Kamenshchik:2020yyn, Kamenshchik:2021tjh, Kamenshchik:2024ybm, Miao:2021gic, Miao2} and references therein. Also, for scalar  field theories with non-derivative interaction, the late time stochastic formalism is an excellent way to do resummation~\cite{Starobinsky:1994bd} (see also e.g.~\cite{Finelli:2008zg, Garbrecht:2013coa, Cho:2015pwa, Ebadi:2023xhq, Cruces:2022imf, Kawasaki:2026hnx} and references therein). However, despite some recent progresses, e.g.~\cite{Kitamoto:2018dek, Miao:2021gic}, the  issue of resummation in  gravity remains largely as an open question to this date.   \\

\noindent
In this work we wish to compute the two loop scalar self energy due to the Yukawa coupling and subsequently, its effect on the loop correction of the coincident two  point correlation function, $\langle \phi^2 \rangle$, in the inflationary de Sitter spacetime. Both the fields will be taken to be massless. Earlier works involve one loop self energy computation, fermion mass generation,  decoherence for the this theory with or without scalar self interaction~\cite{Duffy:2005ue, Prokopec:2003qd,  Garbrecht, Bhattacharya:2023twz, Bhattacharya:2023yhx}. See also~\cite{Elizalde:1993qh, Miao:2006pn, Miao:2020zeh, Boyanovsky:2016exa, Chen:2016nrs, Bhattacharya:2025jtp, Hardwick:2019uex} for  effective action calculation for Yukawa theory in de Sitter.   We also refer our reader to~\cite{Bhattacharya:2025eqf} for a study on the effect of Yukawa coupling on cosmological correlation functions. We further refer our reader to \cite{Toms:2018wpy, Toms:2018oal, Inagaki:1993ya,  Elizalde:1995bm, Inagaki:2015eza},  for discussion on effective action with Yukawa and four fermion interactions in  general curved spacetimes using the Schwinger-DeWitt local expansion technique. \\
 
\begin{figure}[h!]
\begin{center}
  \includegraphics[width=6.5cm]{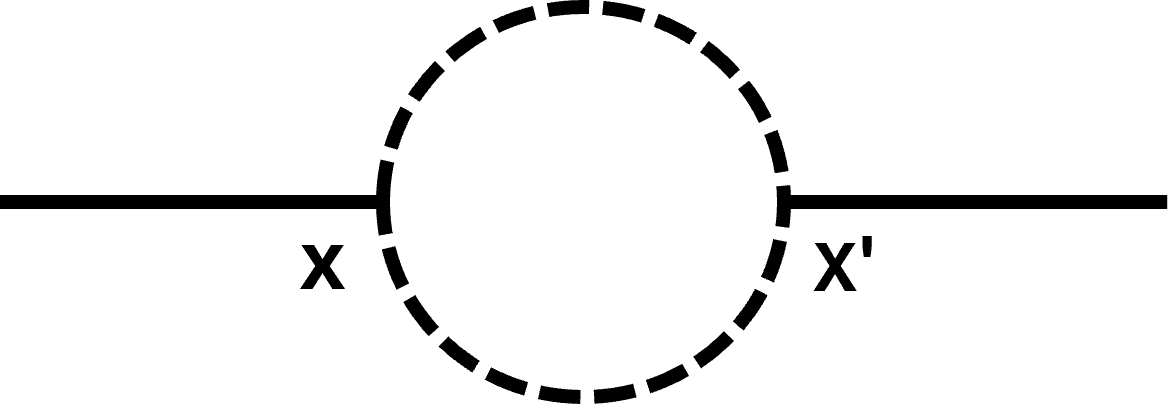} \qquad 
  \includegraphics[width=6.5cm]{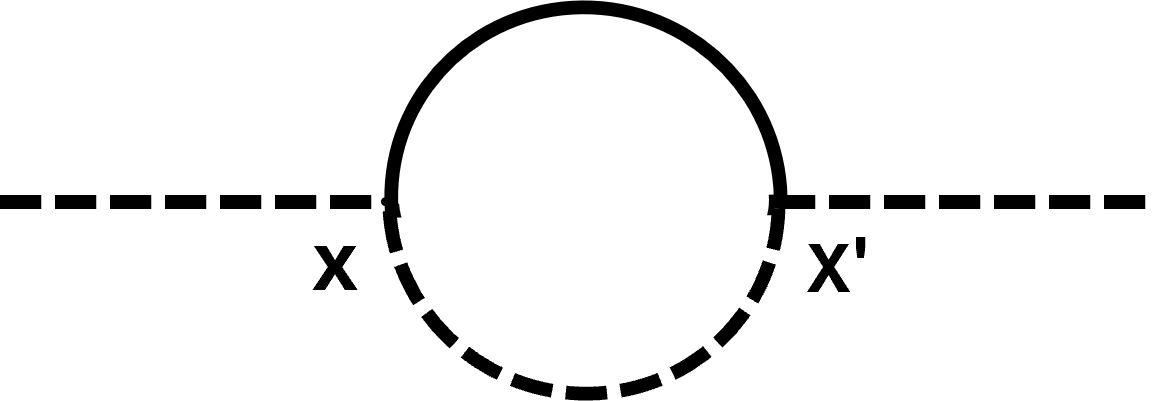}
 \caption{ \it \small One loop scalar and fermion self energy diagrams for the Yukawa interaction. The solid and broken lines respectively represent the scalar and the fermions.}
  \label{f1}
\end{center}
\end{figure}

\noindent
The chief motivation behind this study is as follows. The one loop Yukawa scalar self energy (the first of \ref{f1}) contains a massless or conformal fermion loop. Accordingly the $\ln a$  appearing in the  one loop result, 
 \ref{y14}~\cite{Duffy:2005ue}, is a local or UV secular logarithm, directly originating from renormalisation. The two loop diagrams (\ref{f2}) contain one internal scalar line each and  one expects to obtain IR secular logarithms from them. Nevertheless, we will argue in this paper that this IR logarithms make subleading contributions to the coincident two point correlation function ($\langle \phi^2\rangle$) compared to the local ones in the present context.   We wish to emphasise that the main difference of the present paper with that of the previous ones is going beyond the one loop self energy calculation and subsequent computation of $\langle \phi^2\rangle$, which is resummed afterward. As we have explained, the two loop scalar self energy diagrams  are qualitatively different from the one loop, as the first contains internal scalar lines. A massless fermion is conformally invariant, whereas a massless minimally coupled scalar is not. Apart from this, note  in the amputated diagrams \ref{f2} that not all the internal lines  begin at one common point and converge at another common point, like that of \ref{f1}. This will pose difficulty in evaluating the second two loop diagram directly. To the best of our knowledge, diagrams with such topology have not been attempted earlier in de Sitter. We will also see that we need some non-minimal counterterms to renormalise the two loop divergences -- a feature absent at one loop.  \\

 \noindent
 The rest of this paper is organised as follows. After briefly reviewing the basic setup in the next section we compute after renormalisation, the  UV secular logarithm contribution to the self energy in \ref{2loop}. Next we compute  $\langle \phi^2\rangle$ up to two loop using these self energies. A resummed expression for the same will also be computed numerically. We show that the same has bounded value at late times and it decreases monotonically with the increasing coupling value. Finally we conclude in \ref{fin}.
We will work with the mostly positive signature of the metric in $d=4-\e$ ($\e=0^+$) dimensions  and will set $c=1=\hbar$ throughout.

\section{The basic setup}\label{S2}

We wish to briefly review below the basic framework we will be working in, following chiefly~\cite{Onemli:2002hr, Brunier:2004sb, Miao:2006pn}. 
The metric for the inflationary de Sitter spacetime reads respectively in the cosmological and conformal temporal coordinates
\begin{eqnarray}
ds^2 = -dt^2 + a^2(t) d{\vec x}^2= a^2(\eta) \left(-d\eta^2 +  d{\vec x}^2\right)
\label{y0}
\end{eqnarray}
where $a(t)=e^{Ht}$ or $a(\eta)=  -1/H\eta$ is the de Sitter scale factor and $H=\sqrt{\Lambda/3}$ is the de Sitter Hubble rate, with $\Lambda$ being the cosmological constant.  We have the range $0 \leq t \ensuremath{<} \infty$, so that $-H^{-1}\leq \eta \ensuremath{<}0^-$. Note that any temporal level to the initial hypersurface can 
be a assigned as we wish, by exploiting the time translation symmetry, with a subsequent scaling of the spatial coordinates of the de Sitter.\\ 

\noindent
The bare action corresponding to the matter sector reads,
\begin{eqnarray}
S= \int a^d d^d x \left[ -\frac12 g^{\mu\nu} (\nabla_{\mu} \Phi')(\nabla_{\nu} \Phi')  - \frac{\lambda_0 \Phi'^4}{4!}   + i\overline{\Psi}\slashed{\nabla} \Psi  -g_0 \overline{\Psi} \Psi \Phi'   \right]
\label{y1}
\end{eqnarray}
where $\slashed{\nabla}= \tilde{\gamma}^{\mu}\nabla_{\mu}$, and $\tilde{\gamma}^{\mu}$ and $\nabla_{\mu}$ are respectively the curved space gamma matrices and the spin covariant derivative. Also, since we are working with the mostly positive signature of the metric, we will take the anti-commutation relation for the $\gamma$-matrices as
\be
[\tilde{\gamma}_{\mu},\tilde{\gamma}_{\nu}]_+= -2 g_{\mu\nu} \, {\bf I}_{d\times d}=-2 a^2(\eta)\eta_{\mu\nu} \, {\bf I}_{d\times d}
\label{Ac}
\ee
Thus we may choose, $\tilde{\gamma}_{\mu} = a(\eta)\gamma_{\mu}$, where $\gamma_{\mu}$'s are the flat space gamma matrices.

Defining the field strength renormalisation, $\phi=\Phi'/\sqrt{Z}$ and $\psi=\Psi/\sqrt{Z_f}$, we have
\begin{eqnarray}
&&S= \int a^d d^d x\left[ -\frac{Z}{2} \eta^{\mu\nu} a^{-2} (\p_{\mu} \phi)(\p_{\nu} \phi)   - \frac{Z^2 \lambda_0}{4!} \phi^4     + i Z_f\overline{\psi}\slashed{\nabla} \psi  -g_0 Z_f \sqrt{Z}\, \overline{\psi} \psi \phi\right] 
\label{y2}
\end{eqnarray}
 We next write
\begin{eqnarray}
&&Z= 1+\delta Z, \qquad Z m^2 = \delta m^2, \qquad Z^2 \lambda_0 = \lambda +\delta \lambda \nonumber\\ && 
Z_f =1+\delta Z_f, \qquad M Z_f= \delta M, \qquad g_0 Z_{f}\sqrt{Z}= g+\delta g.  
\label{y3}
\end{eqnarray}
The above decomposition splits \ref{y2} as
\begin{eqnarray}
&&S= \int a^d d^d x\left[  -\frac{1}{2} \eta^{\mu\nu} a^{-2} (\p_{\mu} \phi)(\p_{\nu} \phi)  - \frac{\lambda \phi^4}{4!}       + i \overline{\psi}\slashed{\nabla} \psi    -g \overline{\psi} \psi \phi \right.\nonumber\\&&\left.-\frac{\delta Z}{2} \eta^{\mu\nu} a^{-2} (\p_{\mu} \phi)(\p_{\nu} \phi)  - \frac12 \delta m^2 \phi^2-\frac{\delta \lambda \phi^4}{4!}     + i \delta Z_f \overline{\psi}\slashed{\nabla} \psi  -\delta M \overline{\psi}\psi  - \delta g \overline{\psi} \psi \phi\right]
\label{y4}
\end{eqnarray}

\noindent
Let us first consider the scalar field. A massless and minimally coupled scalar field  is of special interest in de Sitter, owing to its de Sitter symmetry breaking property.  The relevant propagator reads~\cite{Onemli:2002hr},
\be
i\Delta(x,x')= A(x,x')+ B(x,x')+ C(x,x')
\label{props1}
\ee
where 
\begin{eqnarray}
&&A(x,x') = \frac{H^{2-\e} \Gamma(1-\e/2)}{4\pi^{2-\e/2}}\frac{1}{y^{1-\e/2}} = \frac{\Gamma(1-\e/2)}{2^2 \pi^{2-\e/2}}\frac{(aa')^{-1+\e/2}}{(\Delta x)^{2-\e}} \nonumber\\
&&B(x,x') =  \frac{H^{2-\e} \Gamma(3-\e)}{2^{4-\e}\pi^{2-\e/2}\Gamma(2-\e/2)}\left[-\frac{2\Gamma(3-\e/2) \Gamma(2-\e/2)(aa'H^2/4)^{\e/2} }{\e \Gamma(3-\e)}  (\Delta x^2)^{\e/2}+  \left(\ln (aa')+ \frac{2}{\e}\right)  \right]\nonumber\\
&&C(x,x')= \frac{H^{2-\e} }{(4\pi)^{2-\e/2}} \sum_{n=1}^{\infty} \left[\frac{\Gamma(3-\e+n)}{n\Gamma(2-\e/2+n)}\left(\frac{y}{4} \right)^n- \frac{\Gamma(3-\e/2+n)}{(n+\e/2)\Gamma(2+n)}\left(\frac{y}{4} \right)^{n+\e/2}  \right]
\label{props2}
\end{eqnarray}
The de Sitter invariant biscalar interval reads
\be
y(x,x')= aa'H^2 \Delta x^2 = a(\eta)a(\eta')H^2 \left[ |\vec{x}-\vec{x}'|^2- (\eta-\eta')^2\right]
\label{props3}
\ee
The $\ln (aa')$ term appearing in $B(x,x')$  in \ref{props2} makes the propagator non-invariant under de Sitter symmetry transformations. This is a unique feature of an exactly massless and minimal scalar field theory in de Sitter. In other words, a smooth $m^2\to 0$ limit for a scalar in de Sitter may not even exist. 

There are four propagators pertaining to the in-in or the Schwinger-Keldysh formalism one needs to use in a cosmological framework e.g.~\cite{Hu, Berges:2004yj}, characterised by suitable four complex distance functions, $\Delta x^2$,
\begin{eqnarray}
&&\Delta x^2_{++} =\left[ |\vec{x}-\vec{x'}|^2- (|\eta-\eta'|-i\e)^2\right]= (\Delta x^2_{--})^{*}\nonumber\\
&&\Delta x^2_{+-} =\left[ |\vec{x}-\vec{x'}|^2- ((\eta-\eta')+i\e)^2\right]= (\Delta x^2_{-+})^{*} \qquad (\e=0^+)
\label{props3'}
\end{eqnarray}
The first two correspond respectively to the Feynman and anti-Feynman propagators, whereas the last two correspond to the two Wightman functions. 
From \ref{props2}, we have in the coincidence limit for all the four propagators
\be 
i\Delta(x,x) = \frac{H^{2-\e} \Gamma(2-\e)}{2^{2-\e} \pi^{2-\e/2} \Gamma(1-\e/2)}\left(\frac{1}{\e}+\ln a  \right)
\label{y6}
\ee

Using the above expression, we may compute, for example, the one loop self energy bubble corresponding to the quartic self interaction diagrams.   The corresponding  one loop  mass renormalisation counterterm reads~\cite{Brunier:2004sb}
\be 
\delta m_{\lambda}^2 = - \frac{\lambda H^{2-\e} \Gamma(2-\e)}{2^{3-\e} \pi^{2-\e/2} \Gamma(1-\e/2)\e}
\label{y7}
\ee

\noindent
Let us now come to the case of the fermions. The corresponding Lagrangian density is read off from \ref{y4},
$${\cal L}= a^d \left[i\bar{\psi}\tilde{\gamma}^{\mu}\nabla_{\mu}\psi - (g+\delta g)\bar{\psi}\psi \phi\right]$$
where 
$$\nabla_{\mu}\psi =\p_{\mu}\psi + i\Gamma_{\mu}\psi,$$
is the spin covariant derivative with the spin connection matrices, 
$$\Gamma_{\mu}= \frac{i}{8}\omega_{\mu a b}[\gamma^a, \gamma^b ].$$
$\omega_{\mu a b}$ are the Ricci rotation coefficients defined as  
$$\omega_{\mu a b}= e^{\nu}_a\left(\p_{\mu}e_{\nu b}-\Gamma^{\lambda}_{\mu\nu}e_{\lambda b}\right),$$
where $e_{\mu a}$ are the tetrads. The Greek and Latin indices respectively correspond to the spacetime and local Lorentz indices.  For \ref{y0}, we take $e_{\mu}^b = a^{-1}\p_{\mu} x^{b}$. We have
$$\omega_{\mu a b}= Ha \left( \eta_{\mu a}\delta^0_b - \eta_{\mu b}\delta_a^0\right)$$
Also, corresponding to \ref{Ac}, we take $\tilde{\gamma}^{\mu}= a^{-1}\gamma^{\mu}$, where $\gamma^{\mu}$ is the usual flat space gamma matrix. Using these, the fermion  Lagrangian density simplifies to
$${\cal L}=   \left[ia^{\frac{d-1}{2}}\bar{\psi}\gamma^{\mu}\p_{\mu} a^{\frac{d-1}{2}}\psi - a^d(g+\delta g)\bar{\psi}\psi \phi\right]$$
Thus owing to the conformal invariance of the massless fermions, the propagator for $\psi $ reads
\be
iS(x,x')= -\frac{\Gamma(d/2)}{2\pi^{d/2} (aa')^{(d-1)/2}}\frac{i \gamma_{\mu}\Delta x^{\mu}}{\Delta x^{4-\e}}.
\label{propy}
\ee
Different kind of propagators ($S_{\pm \pm}$) in the in-in formalism can be found from the above by using the different distance functions, \ref{props3'}.

\section{Renormalisation of scalar self energy}\label{S3}
We begin by very briefly considering the one loop self energies for the scalar and the fermion, respectively computed in~\cite{Duffy:2005ue} and \cite{Prokopec:2003qd}.
The one loop scalar self energy reads
\begin{eqnarray}
-i \Sigma (x,x')^{\rm 1-loop}_{ss'}= \lambda_{ss'} g^2 (aa')^d {\rm Tr} iS_{ss'}(x,x') i S_{ss'}(x',x), \qquad ({\rm no~sum~on}~s,s')
\label{y8}
\end{eqnarray}
where $s,s'= \pm$, and  $\lambda_{++}= \lambda_{--}=1$ and $\lambda_{+-}= \lambda_{-+}=-1$. Using \ref{A4}, the $++$-self energy reads
\begin{eqnarray}
&&-i \Sigma (x,x')^{\rm 1-loop}_{++}= - \frac{g^2d\Gamma^2(d/2)(aa')}{2^2 \pi^d}\frac{1}{\Delta_{++}^{6-2\e}}
=\frac{i\mu^{-\e}d g^2\Gamma(1-\e/2) aa'}{2^4\pi^{2-\e/2}(1-\e)\e}\p^2 \delta^d (x-x') + \frac{g^2 aa'}{2^5 \pi^4} \p^4 \frac{\ln \mu^2 \Delta x^2_{++}}{\Delta x^2_{++}}.
\label{y9}
\end{eqnarray}
The mixed type self energies are free of divergences, and hence needs no renormalisation
\begin{eqnarray}
&&-i \Sigma (x,x')^{\rm 1-loop}_{+-}=  \frac{g^2d\Gamma^2(d/2)(aa')}{2^2 \pi^d}\frac{1}{\Delta_{+-}^{6-2\e}}
=- \frac{g^2 aa'}{2^5 \pi^4}\p^4 \frac{\ln \mu^2 \Delta x^2_{+-}}{\Delta x^2_{+-}}.
\label{y10}
\end{eqnarray}
The $--$ and $-+$ self energies can be found from the complex conjugation of \ref{y9}, \ref{y10} respectively. 

In order to renormalise the divergence of \ref{y9}, we rewrite the relevant part as 
\begin{eqnarray}
&& - \frac{g^2d\Gamma^2(d/2)(aa')}{2^2 \pi^d}\frac{1}{\Delta_{++}^{6-2\e}}
=\frac{i\mu^{-\e} dg^2\Gamma(1-\e/2) (aa')^{d/2-1}}{2^4\pi^{2-\e/2}(1-\e)\e}\p^2 \delta^d (x-x')+\frac{ig^2}{2^3\pi^{2}} (aa')\ln (aa') \p^2 \delta^d (x-x')
\label{y11}
\end{eqnarray}
One adds with the above contributions from the scalar field strength renormalisation and  non-minimal coupling counterterms
\begin{eqnarray}
&& i\delta Z \p_{\mu}\left( a^{d-2} \p^{\mu}\delta^d (x-x')\right) +i \delta \xi (d-1) a^{d-2} \left( \frac{2\p^2 a}{a}- \e \frac{\eta^{\mu\nu}\p_{\mu}a\p_{\nu}a}{a^2}\right) \delta^d (x-x'),
\label{y12}
\end{eqnarray}
with the choices
\begin{eqnarray}
&& \delta \xi = -\frac{(d-2)}{4}\delta Z, \qquad \delta Z = -\frac{\mu^{-\e}d g^2}{2^4 \pi^{2-\e/2}}\frac{\Gamma(1-\e/2)}{(1-\e)\e}, 
\label{y13}
\end{eqnarray}
so that \ref{y12} becomes 
$$ i\delta Z (aa')^{d/2-1}\p^2 \delta^d (x-x')+ a^d \frac{3H^2 g^2 }{2^3 \pi^2}\delta^d (x-x')$$
The second term  can be cancelled via a further choice of a finite conformal counterterm. Thus the renormalised self energy becomes 
\begin{eqnarray}
&&-i \Sigma (x,x')^{\rm 1-loop, Ren.}_{++}
=\frac{ig^2 (aa')}{2^3\pi^{2}} \ln (aa') \p^2 \delta^4 (x-x')
 + \frac{g^2 aa'}{2^5 \pi^4} \p^4 \frac{\ln \mu^2 \Delta x^2_{++}}{\Delta x^2_{++}}.
\label{y14}
\end{eqnarray}
\bigskip
\noindent
Likewise, the $++$ one loop self energy of the fermion reads
\begin{eqnarray}
&&-i \Sigma_f (x,x')^{\rm 1-loop}_{++}= \frac{g^2 \Gamma(2-\e/2)\Gamma(1-\e/2)(aa')^{(d-1)/2}}{2^3 \pi^{4-\e}} (aa')^{\e/2} \frac{i\gamma_{\mu}\Delta x^{\mu}}{\Delta^{6-2\e}_{++}}
= -\frac{g^2\Gamma^2(1-\e/2) (aa')^{(d-1)/2}}{2^5 \pi^{4-\e}} (aa')^{\e/2} i\slashed{\p}\frac{1}{\Delta x^{4-2\e}_{++}}\nonumber\\ &&=-\frac{g^2\Gamma^2(1-\e/2) (aa')^{(d-1)/2}}{2^5 \pi^{4-\e}} \left( 1+ \frac{\e}{2}\ln (aa')\right) i\slashed{\p} \left(-\frac{2i \mu^{-\e} \pi^{2-\e/2}}{(1-\e)\e \Gamma(1-\e/2)} \delta^d (x-x') -\frac{1}{2^2} \p^2 \frac{\ln \mu^2 \Delta x_{++}^2}{\Delta x_{++}^2}\right)\nonumber\\ &&=
\frac{i\mu^{-\e} g^2 \Gamma(1-\e/2)(aa')^{(d-1)/2}}{2^4 \pi^{2-\e/2}(1-\e)\e}i\slashed{\p}\delta^d (x-x') + \frac{ g^2 (aa')^{3/2}}{2^5 \pi^2}\left(i \ln (aa')i\slashed{\p}\delta^d (x-x')+\frac{1}{4\pi^2} i\slashed{\p}\p^2 \frac{\ln \mu^2 \Delta x^2_{++}}{\Delta x^2_{++}} \right)
\label{y14a}
\end{eqnarray}
The divergence appearing above can accordingly be cancelled by a fermion field strength renormalisation, 
\begin{eqnarray}
\delta Z_f= -\frac{\mu^{-\e}g^2 \Gamma(1-\e/2)}{2^4 \pi^{2-\e/2}(1-\e)\e}.
\label{y14a1}
\end{eqnarray}
%

\subsection{Renormalisation and the local  secular logarithms at two loop}\label{2loop}
%
\begin{figure}[h!]
\begin{center}
  \includegraphics[width=6.5cm]{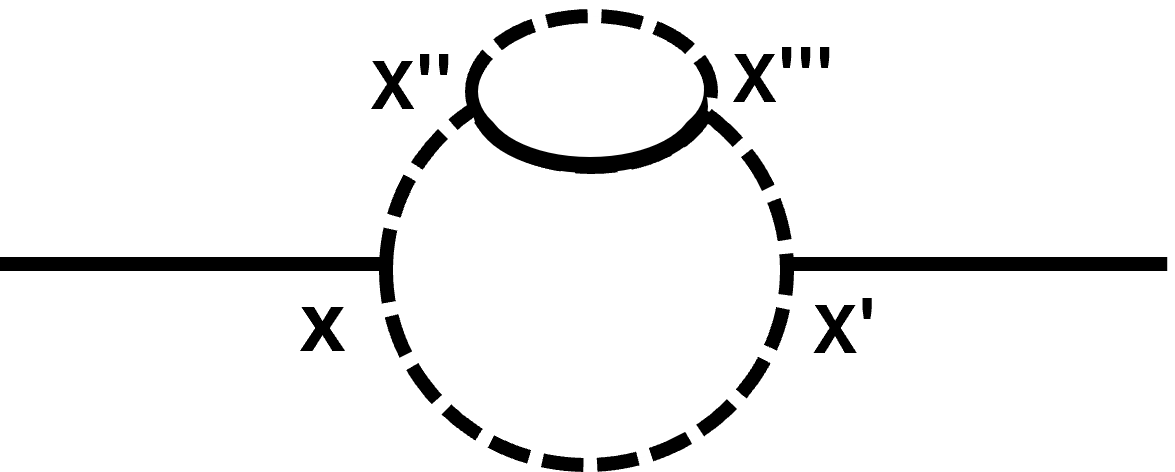}
  \includegraphics[width=6.5cm]{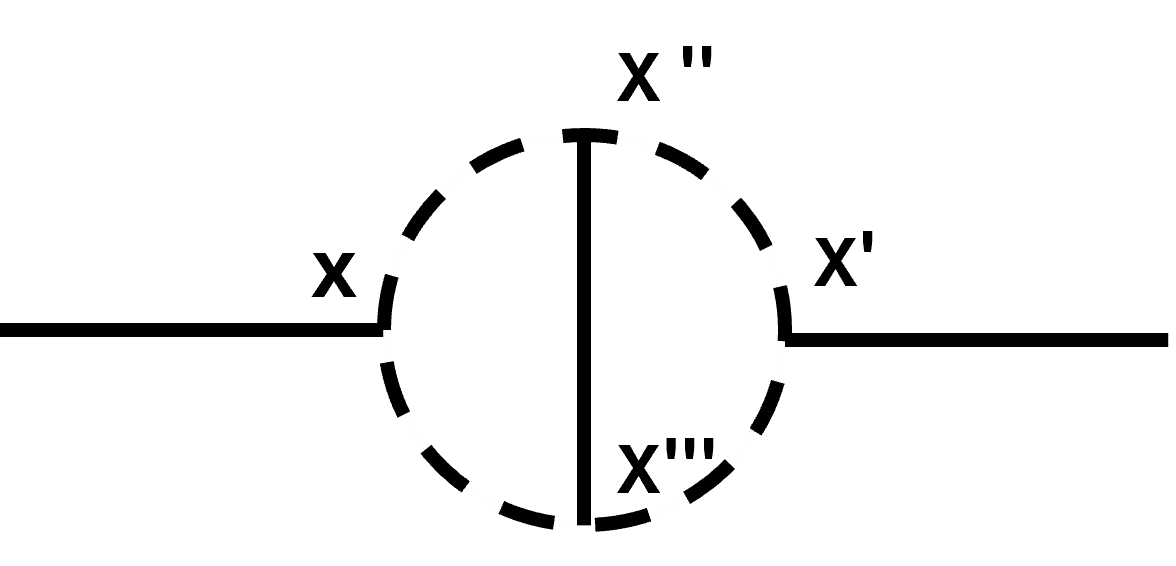}
 \caption{ \it \small Two loop scalar  self energy diagrams for the Yukawa interaction. }
  \label{f2}
\end{center}
\end{figure}
In this section we wish to perform renormalisation and find out the local self energy for the scalar field at two loop.  
There are two two loop diagrams for scalar self energy for the Yukawa interaction, \ref{f2}. Using the basic ingredients described in \ref{S2}, the first reads 
\begin{eqnarray}
&&-i \Sigma_f (x,x')^{\rm 2-loop, 1}_{++}
= -  g^4 (aa')^d{\rm Tr} \int (a''a''')^d d^d x'' d^d x''' iS _{++}(x,x'') iS _{++}(x'',x''')     iS _{++}(x''',x') iS _{++}(x',x) i\Delta_{++}(x'',x''') \nonumber\\ && = -\frac{g^4(aa') \Gamma^4(d/2)}{2^4 \pi^{2d}}\int \frac{(a'' a''')^d d^d x'' d^d x''' }{ (a'' a''')^{d-1}}i\Delta_{++}(x'',x''')   \frac{\Delta x_1^{\mu} \Delta x_2^{\nu} \Delta x_3^{\alpha} \Delta x_4^{\beta}}{\Delta x_{1++}^{4-\e} \Delta x_{2++}^{4-\e} \Delta x_{3++}^{4-\e} \Delta x_{4++}^{4-\e}}  {\rm Tr} \left( \gamma_{\mu}\gamma_{\nu}\gamma_{\alpha}\gamma_{\beta}\right)                                                                                                                                       
\label{y15}
\end{eqnarray}
where the subscripts $1,2,3,4$ respectively stand for the intervals $x-x''$, $x''-x'''$, $x'''-x'$ and $x'-x$. For the purpose of doing renormalisation and finding local part of the self energy, the relevant part of the scalar propagator would be $A(x,x')$, \ref{props2}. This gives 
\begin{eqnarray}
&&-i \Sigma_f (x,x')^{\rm 2-loop, 1}_{++, {\rm loc}}\nonumber\\ &&
 = \frac{g^4 (aa')\Gamma^5(1-\e/2)(1-\e/2)^3}{2^8 \pi^{10-5\e/2}}\int  d^d x'' d^d x'''(a'' a''')^{\e/2}  \left(\p''^{\nu}\frac{1}{ \Delta x_{2++}^{4-2\e} }\right)  \frac{\Delta x_1^{\mu} \Delta x_3^{\alpha} \Delta x_4^{\beta}}{\Delta x_{1++}^{4-\e}\Delta x_{3++}^{4-\e} \Delta x_{4++}^{4-\e}}  {\rm Tr} \left( \gamma_{\mu}\gamma_{\nu}\gamma_{\alpha}\gamma_{\beta}\right)     \Big \vert_{\rm loc}                                                                                                                                  
\label{y16}
\end{eqnarray}
In order to find out the divergence and the local part of the self energy we take the local part of $\Delta x_{++}^{-4+2\e}$ from \ref{A4}, to have 
\begin{eqnarray}
&&-i \Sigma_f (x,x')^{\rm 2-loop, 1}_{++, {\rm loc}}
 = \frac{g^2(aa') \Gamma^3(2-\e/2)}{2^3 \pi^{6-3\e/2}}{\rm Tr} \ \gamma_{\mu}\int d^d x'' d^d x'''  \left(1+ (a''a''')^{\e/2}-1 \right)\left(- \frac{g^2\mu^{-\e}\Gamma(1-\e/2)}{2^4\pi^{2-\e/2}(1-\e)\e}i\slashed{\p}'' \delta^d (x''-x''')\right) \nonumber\\ && \times  \frac{\Delta x_1^{\mu} \Delta x_3^{\alpha} \Delta x_4^{\beta}}{\Delta x_{1++}^{4-\e}\Delta x_{3++}^{4-\e} \Delta x_{4++}^{4-\e}} \gamma_{\alpha}\gamma_{\beta}.                                                                                                                                    
\label{y17}
\end{eqnarray}
We now add with the above the contribution from the fermion field strength renormalisation counterterm, \ref{y14a1},
\begin{eqnarray}
- g^2 \delta Z_f (aa') \frac{\Gamma^3(2-\e/2)}{2^3\pi^{6-3\e/2}}{\rm Tr} \int d^d x'' d^d x''' \frac{\gamma_{\mu}\Delta x_1^{\mu} }{\Delta x_{1++}^{4-\e}} i\slashed{\p}'' \delta^d (x''-x''') \frac{\gamma_{\nu}\Delta x_3^{\nu} }{\Delta x_{3++}^{4-\e}}\frac{\gamma_{\alpha}\Delta x_4^{\alpha} }{\Delta x_{4++}^{4-\e}}                                                                                    
\label{y18}
\end{eqnarray}
where the intervals like $\Delta x_1^{\mu}$ have been defined below \ref{y15}. This precisely cancels the UV divergence of \ref{y17}, to give
\begin{eqnarray}
&&-i \Sigma_f (x,x')^{\rm 2-loop, 1}_{++, {\rm loc}}
 = \frac{i\mu^{-\e} g^4 (aa')\Gamma^4(1-\e/2)}{2^{10}\pi^{8-2\e}(1-\e)\e} {\rm Tr} (\gamma^{\mu}\gamma^{\nu}\gamma^{\alpha}\gamma^{\beta}) \int d^d x'' d^d x'''  \left((a''a''')^{\e/2}-1 \right)\nonumber\\ && \times  \p''_{\nu}\delta^d (x''-x''') \p_{\mu}\frac{1}{\Delta x_{1++}^{2-\e}}  \p'''_{\alpha}\frac{1}{\Delta x_{3++}^{2-\e}}\p'_{\beta}\frac{1}{\Delta x_{4++}^{2-\e}} \Big \vert_{\rm loc} \nonumber\\ &&
%
  = \frac{i\mu^{-\e}d g^4 (aa')(1-\e/4)\Gamma^4(1-\e/2)}{2^{8}\pi^{8-2\e}(1-\e)\e}  \int d^d x'' d^d x'''  \left((a''a''')^{\e/2}-1 \right)\left[ \p''^{\mu}\delta^d (x''-x''')\p_{\mu} \frac{1}{\Delta x_{1++}^{2-\e}} \p'''^{\alpha}\frac{1}{\Delta x_{3++}^{2-\e}}\p'_{\alpha}\frac{1}{\Delta x_{4++}^{2-\e}}\right. \nonumber\\ && \left. -\p''^{\mu}\delta^d (x''-x''')\p'_{\mu} \frac{1}{\Delta x_{4++}^{2-\e}} \p'''^{\alpha}\frac{1}{\Delta x_{3++}^{2-\e}}\p_{\alpha}\frac{1}{\Delta x_{1++}^{2-\e}}  + \p''^{\mu}\delta^d (x''-x''')\p'''_{\mu} \frac{1}{\Delta x_{3++}^{2-\e}} \p^{\alpha}\frac{1}{\Delta x_{1++}^{2-\e}}\p'_{\alpha}\frac{1}{\Delta x_{4++}^{2-\e}} \right]_{\rm loc}                                                                                                   
\label{y19}
\end{eqnarray}
Note that in the flat spacetime ($a=1$), the above amplitude vanishes. This means that only non-local contribution is possible for this diagram in the flat spacetime. In order to see  what happens in de Sitter, we integrate \ref{y17} by parts. Noting that both $\p^2$ and $\p'_{\mu}\p^{\mu}$ acting on $\Delta x^{-2+\e}(x,x')$ give $\delta^d(x-x')$ but with opposite signs, it is easy to see that \ref{y17} has no {\it local} parts, i.e., terms  proportional to $\delta^d (x-x')$. 
\begin{figure}[h!]
\begin{center}
  \includegraphics[width=4.5cm]{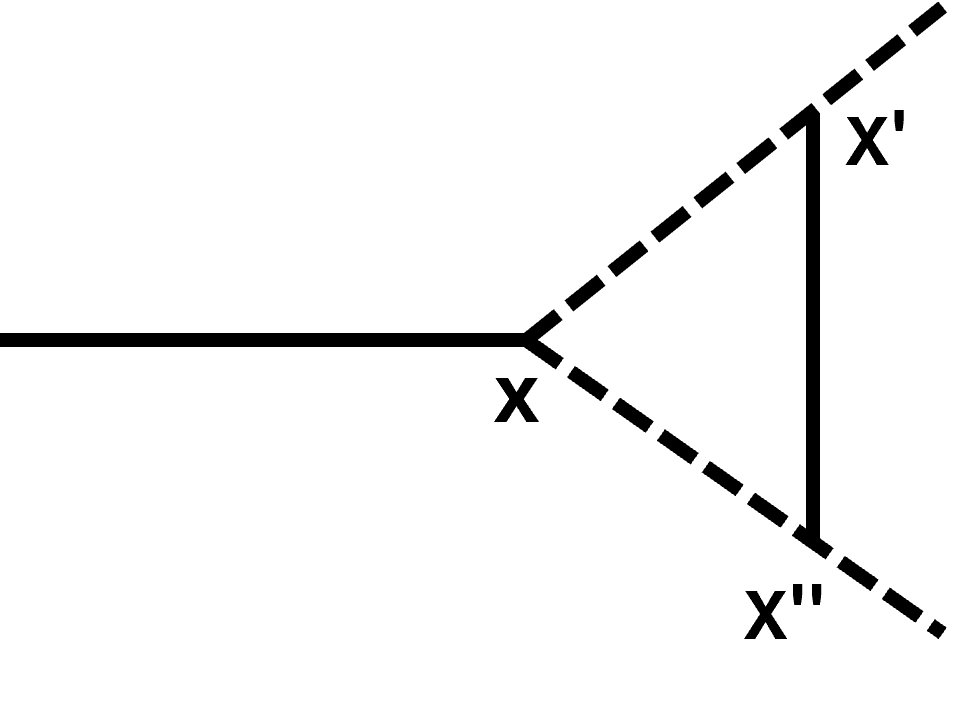}
 \caption{ \it \small One loop vertex diagram for the Yukawa interaction. }
\end{center}
\end{figure}

\noindent
Let us now come to the second diagram, 
\begin{eqnarray}
&&-i \Sigma_f (x,x')^{\rm 2-loop, 2}_{++}
=  g^4 (aa')^d{\rm Tr} \int (a''a''')^d d^d x'' d^d x''' iS _{++}(x,x'') iS _{++}(x'',x')     iS _{++}(x',x''') iS _{++}(x''',x) i\Delta_{++}(x'',x''') \nonumber\\ &&
=     \frac{g^4 d \Gamma^4(2-\e/2) (aa')}{2^4 \pi^{8-2\e}} \int  a'' a''' d^d x'' d^d x''' i\Delta_{++}(x'',x''')  \left[ \frac{-i\Delta x_{1}^{\mu}}{\Delta x_{1++}^{4-\e}}\frac{-i\Delta x_{4\mu}}{\Delta x_{4++}^{4-\e}}  \frac{-i\Delta x_2^{\nu}}{\Delta x_{2++}^{4-\e}} \ \frac{-i\Delta x_{3\nu}}{\Delta x_{3++}^{4-\e}} + (1, \ 2~{\rm and}~3, \ 4~{\rm contractions})\right. \nonumber\\ &&
\left. - (1,\ 3~{\rm and}~2, \ 4~{\rm contractions})\right],                                                                                                       
\label{y20}
\end{eqnarray}
where the subscripts $1,2,3,4$ respectively denote the intervals $x-x''$, $x''-x'$, $x'-x'''$ and $x'''-x$. This seems to be a very complicated diagram to deal with in the coordinate space. The reason behind this is the product of distance functions between four different points. One way to deal with this could be to introduce the Feynman parameters, define a new coordinate and then use \ref{A4}, in order to create distance functions like $\sim (x-x_0)^{-\alpha}$ ($\alpha\ensuremath{>}0$). Such new coordinates will be the linear combination of the old coordinates. However in doing so, we face an immediate problem of breaking the de Sitter or even the FRW invariance due to the appearance of the scale factor in the amplitudes. As of now, it remains elusive to us how we may deal with this integral directly.

In order to bypass this issue, and to do renormalisation, we will attempt to use an analogue of a relationship called the Donoghue identity~\cite{Donoghue:1993eb, Donoghue:1994dn}, which  in flat and de Sitter spacetime reads~\cite{Woodard:2025smz},
$$i\Delta_{m++} (x,x')i\Delta_{++}(x,x'')i\Delta_{++}(x',x'') \longrightarrow \frac{i\delta^d (x-x')}{2m^2}i\Delta_{++}^2(x,x''),$$
The above relationship  may help us to determine the local part (and hence the renormalisation) as well as some non-local part of the respective amplitude. In the above expression $i\Delta_{m}$ corresponds to a massive scalar, whereas the rest are massless propagators. We wish to find out something analogous to the above  for the Yukawa theory, which for our case contains only massless propagators.

To proceed, let us first recall the one loop vertex function for the Yukawa theory in the {\it flat} spacetime. The local or divergent part reads
\begin{eqnarray}
&&-i V^{\rm 1-loop, div.} = g^3 \int \frac{d^d l}{(2\pi)^d} \frac{\slashed{l}+\slashed{p}_1}{(l+p_1)^2} \frac{\slashed{l}+\slashed{p}_2}{(l+p_2)^2}\frac{1}{l^2}\Big\vert_{\rm loc.}   \to -g^3  \int \frac{d^d l}{(2\pi)^d} \frac{1}{ l^2 (l+p_1-p_2)^2 }\Big\vert_{\rm loc.}                                                                                    
\label{y21}
\end{eqnarray}
Thus the divergent and hence the local part of the integral can simply be thought of as a product of two massless scalar propagators, with the same starting and end points.  Keeping in mind the relationship dictated by the Donoghue identity,
 we now analogously write down the correspondence for the {\it local} part of the Yukawa vertex function in de Sitter in coordinate space,
\begin{eqnarray}
&&-i a^d V^{\rm 1-loop} (x,x'',x''')\Big\vert_ {\rm loc.}= -i g^3 (aa''a''')^d {\rm Tr}\ iS_{++}(x,x'')iS_{++}(x''',x) i\Delta_{++}(x'',x''')\vert_{\rm loc.} \nonumber\\ &&
\longrightarrow  - a^d g^3  \delta^d(x-x'') i\Delta^2_{++}(x'',x''')\vert_{\rm loc.}= \frac{i\mu^{-\e}  g^3 \Gamma(1-\e/2)}{2^3 \pi^{2-\e/2}(1-\e)\e} a^d \left(1+\e\ln a +\frac{\e^2}{2}\ln^2 a +{\cal O}(\e^3)\right) \delta^d (x-x'')\delta^d (x''-x''').\nonumber\\
\label{y22}
\end{eqnarray}
Note that the first expression for the second line contains product of two massless scalar propagators, as of \ref{y21}. While the original one loop Yukawa vertex contains only one such propagator, we emphasise that the above correspondence has to be understood in terms of the negative powers of the distance functions ($\Delta x^2_{++}$) appearing in the scalar and fermion propagators.\\  

\ref{y22} gives us the one loop Yukawa vertex renormalisation counterterm,
$$\delta g_{\rm 1-loop}=\frac{\mu^{-\e} g^3 \Gamma(1-\e/2)}{2^3 \pi^{2-\e/2}(1-\e)\e} $$
Thus the local part of the  one loop renormalised Yukawa vertex function in de Sitter spacetime reads 
\begin{eqnarray}
&&-i V^{\rm 1-loop} (x,x'',x''')\Big\vert_ {\rm loc., Ren.}= \frac{i g^3}{8 \pi^{2}} \left(\ln a +\frac{\e}{2}\ln^2 a\right) \delta^4 (x-x'')\delta^4 (x''-x''') +{\cal O}(\e^2)
\label{y23}
\end{eqnarray}
Note that this vanishes in the flat spacetime, $a=1$. The non-vanishing finite local vertex function will create some obstacles in the renormalisation of the second of  \ref{f2}, as we will see below. 

From \ref{y20}, it is easy to see that only the first contraction of the second line can give a local, or divergent contribution. We have
\begin{eqnarray}
&&-i \Sigma_f (x,x')^{\rm 2-loop, 2}_{++, \rm loc.}
=     \frac{g^4 d \Gamma^2(2-\e/2) a^{1+\e}a'}{2^2 \pi^{4-\e}} \cdot \frac{\mu^{-\e}\Gamma(1-\e/2)}{2^3 \pi^{2-\e/2}(1-\e)\e}   \   \frac{1}{\Delta x^{6-2\e}_{++}} =  \frac{\mu^{-\e} d g^4 \Gamma(1-\e/2)\Gamma^2 (2-\e/2)}{2^5 \pi^{6-3\e/2}(1-\e)\e}(aa')^{1-\e/2} \frac{1}{\Delta x^{6-2\e}_{++}}(aa')^{\e}
 \nonumber\\&&    
\label{y24}
\end{eqnarray}
 The one loop vertex renormalisation counterterm's contribution reads
\begin{eqnarray}
&&-i \Sigma_f (x,x')^{\delta g_{\rm 1-loop}}_{++}= - \frac{\mu^{-\e} d g^4 \Gamma(1-\e/2)\Gamma^2 (2-\e/2)}{2^5 \pi^{6-3\e/2}(1-\e)\e}(aa')^{1-\e/2} \frac{1}{\Delta^{6-2\e}_{++}}(aa')^{\e/2}
\label{y25}
\end{eqnarray}
Adding the above with \ref{y24}, we have,
\begin{eqnarray}
&&-i \Sigma_f (x,x')^{\rm 2-loop, 2}_{++}+ -i \Sigma (x,x')^{\delta g_{\rm 1-loop}}_{++}
=  \frac{\mu^{-\e} d g^4 \Gamma(1-\e/2)\Gamma^2 (2-\e/2)}{2^6 \pi^{6-3\e/2}(1-\e)}(aa')^{1-\e/2} \frac{1}{\Delta^{6-2\e}_{++}}\left(\ln (aa') +\frac{3\e}{4} \ln^2(aa')\right) +{\cal O}(\e^2)
 \nonumber\\&&    =
 - \frac{i\mu^{-2\e} dg^4 \Gamma^2(1-\e/2)}{2^7 \pi^{4-\e}(1-\e)^2}(aa')^{1-\e/2} \frac{\ln a }{\e} \p^2 \delta^d (x-x') - \frac{3i g^4}{2^6 \pi^4} a^2\ln^2 a \p^2 \delta^4 (x-x') +{\cal O}(\e) + {\rm non-local~terms},
\label{y26'}
\end{eqnarray}
where in the last line we have used  \ref{A4}.
Note that  in the flat spacetime ($a=1$), \ref{y26} shows that the second of \ref{f2} has no renormalised local contribution, which, as we see,  is clearly not the case in de Sitter. In fact the  divergence in \ref{y26'} is problematic, as to the best of our knowledge, there seems to be no traditional counterterm to absorb it.  Accordingly, we introduce  non-canonical counterterms,
\be
\frac13 \int  d^d x a^d \ \delta \zeta^{(1)} \phi^3 \Box \phi + \frac{1}{12} \int  d^d x a^d \ \left(\delta \zeta^{(2)}+\delta \zeta^{(3)} \right) R \phi^4 
\label{y27-}
\ee
The corresponding contribution to the self energy reads
\begin{eqnarray}
&&i \delta \zeta^{(1)} i\Delta(x,x) \p_{\mu} a^{d-2} \p^{\mu}\delta^d (x-x') + i \delta \zeta^{(1)} \left[ \p_{\mu}\left(a^{d-2}\p^{\mu}i\Delta(x,x) \right)- 2a^{d-2} \p_{\mu}\p^{\mu} i\Delta(x,x) \right]\delta^d (x-x') \nonumber\\&& +i a^d R \left(\delta\zeta^{(2)}+  \delta \zeta^{(3)}\right)   i\Delta(x,x) \delta^d (x-x')  \nonumber\\&&    =
i\delta \zeta^{(1)} C \left(\frac{1}{\e}+\ln a \right)\p_{\mu} a^{d-2}\p^{\mu} \delta^d (x-x') -i\delta \zeta^{(1)} CH^2 (d-3) a^d\delta^d (x-x')+ ia^d \left(\delta\zeta^{(2)}+  \delta \zeta^{(3)}\right)  C R \left(\frac{1}{\e}+\ln a \right)\delta^d (x-x') ,\nonumber\\
\label{y27}
\end{eqnarray}
where we have abbreviated
$$C= \frac{H^{2-\e}\Gamma(2-\e)}{2^{2-\e}\pi^{2-\e/2}\Gamma(1-\e/2)}$$
We add \ref{y27} with \ref{y26}. Using 
$$R= d(d-1)H^2,$$
the middle term in the last line of \ref{y27} can be absorbed via a mass renormalisation counterterm, 
\be
\delta m^2_{\rm 2-loop}= -CH^2 (d-3) \delta \zeta^{(1)} + \frac{CH^2d(d-1) }{\e} \delta \zeta^{(2)},
\label{ct1}
\ee
whereas the very first divergence in the last line of \ref{y27} can be cancelled via a scalar field strength renormalisation counterterm
\be
\delta Z^{\rm 2-loop}= \frac{\delta \zeta^{(1)} C}{\e}
\label{ct2}
\ee

As in the one loop case described at the beginning of \ref{S2}, we choose in \ref{y27}
$$\delta \zeta^{(2)}= - \frac{(d-2)}{4}\delta \zeta^{(1)},$$
so that the choice 
\be
\delta \zeta^{(2)}= \left(\frac{\mu}{H} \right)^{-\e}\frac{\mu^{-\e} H^{-2} d g^4 \Gamma^3(1-\e/2)}{2^{5+\e}\pi^{2-\e/2}\Gamma(1-\e)(1-\e)^3},
\label{ct2}
\ee
removes the $\ln a/\e$ divergence of \ref{y26}.  

In order to fix $\delta \zeta^{(3)}$ in \ref{y27}, we note that
we still need to consider the contribution coming from the ${\cal O}(g^4)$ scalar quartic vertex counterterm that arises from the renormalisation of the ladder diagram,  
$$\delta \lambda = - \frac{2\mu^{-\e}g^4 \Gamma(1-\e/2)(1-\e/4)}{\pi^{2-\e/2} (1-\e)\e} $$
The contribution from $\delta \lambda$ to the one loop scalar self energy reads, 
\begin{eqnarray}
&&-i \Sigma_f (x,x')^{\delta\lambda}_{++}
= -\frac{i \delta \lambda}{2} a^d i\Delta (x,x)\delta^d (x-x') = \frac{i\mu^{-\e}H^{2-\e}g^4(1-\e/4)\Gamma(1-\e)}{2^{2-\e}\pi^{4-\e}\e} \left( \frac{1}{\e}+\ln a\right) a^d \delta^d (x-x')
\label{y26}
\end{eqnarray}
This leads to the choice 
\be
\delta \zeta^{(3)}= - \frac{\mu^{-\e} H^{-2} g^4\Gamma(1-\e/2)}{2^2 \pi^{2-\e/2}(1-\e)(3-\e)\e},
\label{y27}
\ee
in \ref{y27}, thereby completely cancelling \ref{y26}. \\

\noindent
Putting things together now, the local part of the  renormalised two loop scalar self energy for the Yukawa coupling is given by
\be
-i \Sigma_f (x,x')^{\rm 2-loop, Ren.}_{++, {\rm loc}}= - \frac{3ig^4}{2^6 \pi^4} a^2 \ln^2 a \p^2 \delta^4(x-x').
\label{y28}
\ee

\noindent
Before we end this section, we wish to point out that seemingly a more natural approach to renormalise \ref{y26'} could have been to add with it the hybrid contribution from the one loop field strength counterterm, \ref{y13}, and a {\it finite} quartic vertex counterterm,
$$ \delta Z \delta \lambda^{\rm fin.} \int d^d x' (aa')^d i\Delta_{++}(x,x')\Box i\Delta_{++}(x,x')= \frac{i \delta Z \delta \lambda^{\rm fin.}H^{2-\e}\Gamma(2-\e)}{2^{2-\e}\pi^{2-\e/2}\Gamma(1-\e/2)} a^d \left( \frac{1}{\e}+ \ln a\right)$$
Clearly, due to the difference of the  power of the scale factor appearing above and in \ref{y26'} and above, $\delta\lambda^{\rm fin.}$ will not be able to cancel the divergence. 

Finally,  note  also the appearances of the curvature scale, i.e. the de Sitter Hubble rate $H$, in  the (dimensionful) renormalisation counterterms appearing above. The spacetime curvature affects the propagation of the field, and thereby induces a scale in the loop integrals through the propagators, resulting in the appearance of $H$ in various such counterterms and amplitudes. Thus even though our theory is massless, the background curvature {\it in a sense } plays the role of field rest mass.  An earlier known such example  is the one loop bubble self energy for a massless minimally coupled scalar with a quartic self interaction in de Sitter, which requires a mass renormalisation counterterm~\cite{Brunier:2004sb},
$$ \delta m^2 = - \frac{\lambda H^{2-\e}\Gamma(2-\e)}{2^{3-\e}\pi^{2-\e/2}\Gamma(1-\e/2)\e}\cdot$$
Such counterterm is absent in the flat spacetime for a massless scalar. Perhaps the most celebrated example of induction of a curvature scale in a quantum field theory  is the trace and chiral anomalies. We further refer our reader to e.g.~\cite{Bhattacharya:2023yhx, Bhattacharya:2025jtp} and references therein for some similar examples. We also note that this feature will lead to the appearance of curvature scale in the $\beta$-functions and the renormalisation group equations, and may thus reveal important aspect of the asymptotic  behaviour of various couplings. A detailed analysis of this in particular, in the context of gravitons seems to be very interesting.  This completes our discussion on renormalisation.

We wish to show below that the  ultraviolet or local part of the self energy  computed above makes the leading contribution to the coincident two point correlation function, $\langle \phi^2\rangle$, instead of the non-local part. For simplicity of computation, we will use the IR effective framework, along with the already derived expressions for the self energies.

\section{Computation of $\langle \phi^2\rangle$, leading secular behaviour and resummation}\label{NL-self}
%
\begin{figure}[h!]
\begin{center}
  \includegraphics[width=3.5cm]{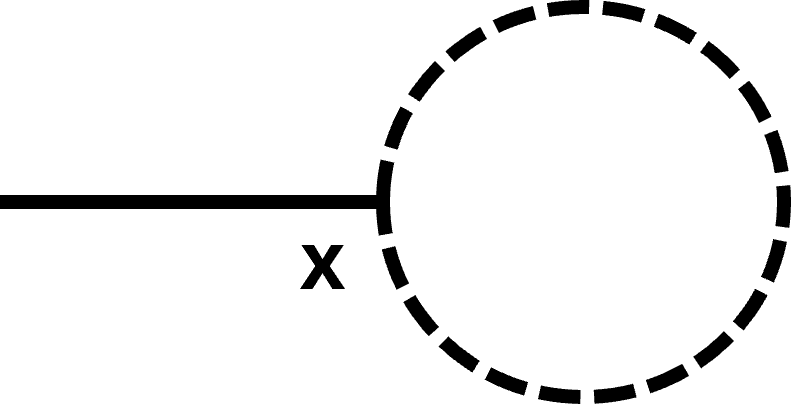} \qquad \qquad \qquad 
  \includegraphics[width=3.0cm]{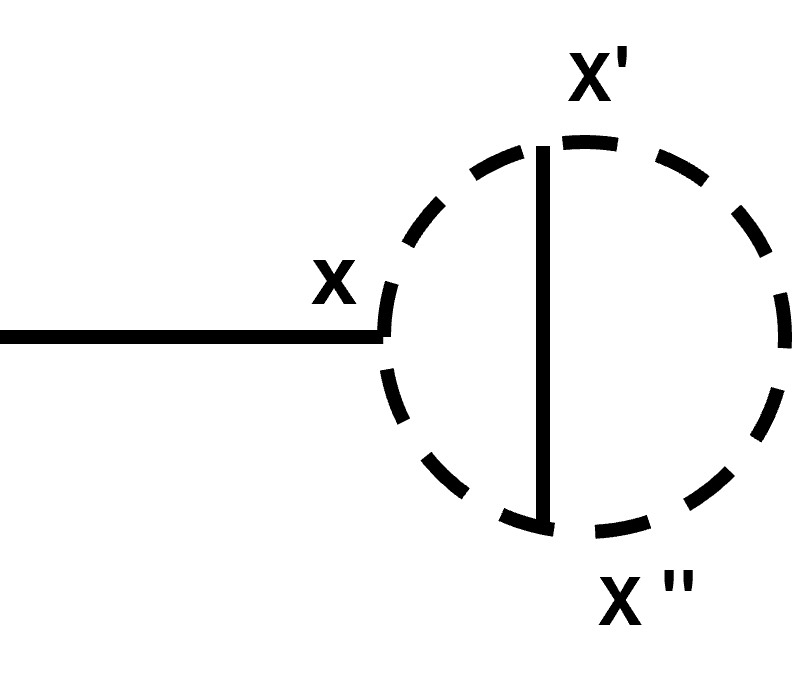}
 \caption{ \it \small The tadpoles at any order of the perturbation theory vanish for the Yukawa theory carrying a massless fermion, for it always involves traces of odd number of $\gamma$-matrices. The primed vertices appearing above are integrated.}
  \label{f3}
\end{center}
\end{figure}
\noindent
In this section we wish to compute the coincident correlation function, $\langle \phi^2(x)\rangle$, owing to the loop effects due to the Yukawa coupling. 
We shall use below the in-in formalism to obtain causal result. Also, instead of using the exact propagators as of the preceding discussion, we will use below an IR effective formalism in three-momentum space, appropriate to capture the leading secular logarithms.  Also note that there is no condensate for the scalar, $\langle \phi \rangle$, via the tadpole diagrams (\ref{f3}) in this case. This is because at any order this involves the trace of odd number of gamma matrices. In order to have a non-zero $\langle \phi \rangle$ with a massless fermion, we must include in the action a cubic self interaction term for the scalar~\cite{Bhattacharya:2023yhx}. 

Writing now $\phi= u_k(\eta) e^{i\vec{k}\cdot \vec{x}}$ ($k= |\vec{k}|$), the normalised mode functions for the massless minimal scalar field reads,
$$u_k (\eta)= \frac{H\eta}{\sqrt{2k^3}}\left(1- \frac{i}{k\eta} \right)e^{-ik\eta}.$$
The field quantisation in the deep IR reads
\begin{eqnarray}
&&\phi(\eta, \vec{x})_{\rm IR}= \int \frac{d^3 \vec{k}}{(2\pi)^{3/2}} \theta (Ha-k)\left[ a_{\vec k} u_k(\eta)_{\rm IR} e^{-i \vec{k}\cdot \vec{x}}+ a^{\dagger}_{\vec k} u^{\star}_k(\eta)_{\rm IR} e^{i \vec{k}\cdot \vec{x}}\right]
\label{y29}
\end{eqnarray}
where $[a_{\vec k}, a^{\dagger}_{\vec k'}]= \delta^3(\vec{k}-\vec{k}')$, and the long wavelength mode functions read,
$$u_k(\eta)_{\rm IR}= \frac{H}{\sqrt{2k^3}} \left[ 1+ \frac12 \left( \frac{k}{Ha}\right)^2 + \frac{i}{3}\left( \frac{k}{Ha}\right)^3 + \cdots \right]$$
The Feynman and anti-Feynman Green functions respectively read
\begin{eqnarray}
&&i\Delta_{++}(k, \eta, \eta')= \theta(\eta-\eta')i\Delta_{-+}(k, \eta, \eta')+ \theta(\eta'-\eta)i\Delta_{+-}(k, \eta, \eta')\nonumber\\&&
i\Delta_{--}(k, \eta, \eta')= \theta(\eta-\eta')i\Delta_{+-}(k, \eta, \eta')+ \theta(\eta'-\eta)i\Delta_{-+}(k, \eta, \eta'),
\label{y30}
\end{eqnarray}
where $i\Delta_{\mp \pm}$ are the positive and negative frequency Wightman functions respectively reading 
\begin{eqnarray}
&&i\Delta_{-+}(k, \eta, \eta')= \frac{H^2}{2k^3}\theta(Ha-k)\theta(Ha'-k)(1+ik\eta)(1-ik\eta') e^{-ik (\eta-\eta')}=\left(i\Delta_{+-}(k, \eta, \eta')\right)^{\star}
\label{y31}
\end{eqnarray}
Thus $\left(i\Delta_{++}(k,\eta,\eta')\right)^{\star}=i\Delta_{--}(k,\eta,\eta') $.

The leading IR behaviour of the Wightman functions are given by
\begin{eqnarray}
i\Delta_{+-}(k, \eta, \eta') && \approx  \frac{H^2}{2k^3}\theta(Ha-k)\theta(Ha'-k)\left( 1+ \frac{ik^3}{3Ha'^3}\right) \qquad (\eta \gtrsim \eta')\nonumber\\ &&
\approx  \frac{H^2}{2k^3}\theta(Ha-k)\theta(Ha'-k)\left( 1- \frac{ik^3}{3Ha'^3}\right) \qquad (\eta' \gtrsim \eta)
\label{y31a1}
\end{eqnarray}
This gives  for $\eta \gtrsim \eta'$,
\begin{eqnarray}
&& i\Delta_{+-}(k, \eta, \eta') - i\Delta_{-+}(k, \eta, \eta')  \approx  \frac{i\theta(Ha -k)\theta(Ha'-k) }{3Ha'^3}, \nonumber\\ && i\Delta_{+-}(k, \eta, \eta') + i\Delta_{-+}(k, \eta, \eta') \approx  \frac{H^2\theta(Ha -k)\theta(Ha'-k) }{k^3}.
\label{y31a2}
\end{eqnarray}
\bigskip

\noindent
Likewise for the fermions, we take the propagators as~\cite{Chen:2016nrs}
\begin{eqnarray}
&& iS_{-+}(k,\eta, \eta')= -\frac{1}{2(aa')^{(d-1)/2}}\frac{i\slashed{k}}{k}e^{-ik(\eta-\eta')}, \quad iS_{+-}(\bar{k},\eta, \eta')=\frac{1}{2(aa')^{(d-1)/2}}\frac{i\slashed{\bar{k}}}{k}e^{ik(\eta-\eta')}\nonumber\\ &&
iS_{++}(k,\bar{k},\eta, \eta')=\theta(\eta-\eta') iS_{-+}(k,\eta, \eta') + \theta(\eta'-\eta)iS_{+-}(\bar{k},\eta, \eta')\nonumber\\ &&
iS_{--}(k,\bar{k},\eta, \eta')=\theta(\eta-\eta') iS_{+-}(\bar{k},\eta, \eta') + \theta(\eta'-\eta)iS_{-+}(k,\eta, \eta')
\label{y32}
\end{eqnarray}
where $\bar{k}\equiv (k^0,-\vec{k})$. Thus  $(iS_{-+})^{\dagger}=iS_{+-}$ and $(iS_{++})^{\dagger}=iS_{--}$.  The deep IR effective limit of \ref{y30}, \ref{y31}, \ref{y32} can be found via the expansion of $k/Ha \lesssim 1$.


Being equipped with these, let us compute $\langle \phi^2 \rangle$ due to one loop $({\cal O}(g^2))$ fermion self energy insertion. For the free theory, the renormalised result is given by~\ref{y7}
$$\langle \phi^2 \rangle_{\rm free} = \frac{H^2}{4\pi} \ln a.$$
We have at one loop
\begin{eqnarray}
&&\langle \phi^2 \rangle_{g^2} = g^2 {\rm Tr}\int (a'a'')^dd^dx' d^d x'' \left[ iS_{++}(x',x'')iS_{++}(x'',x') i\Delta_{++}(x,x')i\Delta_{++}(x,x'')\right.\nonumber\\ &&\left. - iS_{+-}(x',x'')iS_{-+}(x'',x') i\Delta_{++}(x,x')i\Delta_{+-}(x,x'')- iS_{-+}(x',x'')iS_{+-}(x'',x') i\Delta_{+-}(x,x')i\Delta_{++}(x,x'')\right. \nonumber\\ && \left. +iS_{--}(x',x'')iS_{--}(x'',x') i\Delta_{+-}(x,x')i\Delta_{+-}(x,x'') \right]  
\label{y33}
\end{eqnarray}
Let us first compute the correlation due to the fermion local self energy, i.e., the first term on the right hand side of~ \ref{y14}. Note that we will take temporal coordinate  $\eta$ corresponding to $x$ to be the final time, $\eta \gtrsim \eta'$, $\eta \gtrsim \eta''$.

Using \ref{y31a2}, we have for the local self energy
\begin{eqnarray}
&&\langle \phi^2 \rangle_{g^2, {\rm loc}} = \frac{ig^2}{4\pi^2} \int d^4x' d^4 x'' (a'a'')\ln a''  \p''^2\delta^4 (x'-x'')\left[i\Delta_{-+}(x,x')i\Delta_{-+}(x,x'')- i\Delta_{+-}(x,x')i\Delta_{+-}(x,x'') \right]  \nonumber\\ &&
=\frac{ig^2}{4\pi^2} \int d^4x'  a'^2\ln a'\p'^2 \left[  \left(i\Delta^2_{-+}(x,x')- i\Delta^2_{+-}(x,x')\right) \right]=- \frac{g^2H^2 }{12\pi^4}\ln^3 a +\ {\rm subleading~terms}
\label{y34}
\end{eqnarray}
Let us now compute $\langle \phi^2 \rangle_{g^2}$ for non-local self energy. The integral of \ref{y33} can be evaluated for two temporal hierarchies, $\eta'\gtrsim \eta''$ and  $\eta''\gtrsim \eta'$. For the first, we have 
\begin{eqnarray}
&&\langle \phi^2 \rangle_{g^2, {\rm NL}} = g^2 {\rm Tr}\int^{\eta}_{\eta'=-H^{-1}}a'^4 d\eta' d^3 \vec{x}'\int^{\eta'}_{\eta'' =-H^{-1}}a''^4 d\eta'' d^3 \vec{x}'' \ \left( i\Delta_{-+}(x,x')-i\Delta_{+-}(x,x')\right)\nonumber\\ && \times 
\left[ iS_{-+}(x',x'')iS_{+-}(x'',x') i\Delta_{-+}(x,x'') - {\rm h.c.} \right]  \nonumber\\ &&
=\frac{g^2}{H^2} {\rm Tr}\int^{a}_{a'=1}a'^2 da'\int^{a'}_{a'' =1}a''^2 da'' \int \frac{ d^3 \vec{k}_1 d^3 \vec{k}_2}{(2\pi)^6} \left( i\Delta_{-+}(k_1,\eta,\eta')-i\Delta_{+-}(k_1,\eta,\eta')\right)\nonumber\\ &&
\times \left[ iS_{-+}(k_2,\eta',\eta'')iS_{+-}(\bar{l}, \eta'',\eta') i\Delta_{-+}(k_1,\eta,\eta'')- {\rm h.c.}\right]\nonumber\\ &&
=\frac{g^2}{H^2} \int^{a}_{a'=1}\frac{da'}{a'}\int^{a'}_{a'' =1}\frac{da''}{a''} \int \frac{ d^3 \vec{k}_1 d^3 \vec{k}_2}{(2\pi)^6} \frac{\theta(Ha'-k_1)\theta(Ha''-k_1)}{3Ha'^3}\nonumber\\ &&
\times \left( 1+ \frac{\vec{k}_2 \cdot \vec{l}}{l k_2} \ \right)\left[ \frac{1}{3Ha''^3}\cos (l+k_2)(\eta'-\eta'') +\frac{H^2}{k_1^3}\sin (l+k_2)(\eta'-\eta'')  \right]\nonumber\\ && \sim  g^2 H^2 \ln^2 a + {\rm subleading~terms},
\label{y35}
\end{eqnarray}
where we  abbreviated $l=|\vec{k}_2-\vec{k}_1|$ above. The integral involving spacelike inner product vanishes, for it involves odd function. The non-local contribution is subleading compared to that of the local, \ref{y34}. This is not surprising, as there is no internal scalar lines present in this case. The other temporal hierarchy gives the same result as above. \\

\noindent
Let us now consider the two loop case.  The local self energy contribution, \ref{y28}, reads 
\begin{eqnarray}
&&\langle \phi^2 \rangle_{g^4, {\rm loc}} = g^4 \int d^4 x' d^4 x'' d^4 x''' d^4 x'''' (a'a''a'''a'''')^4 \nonumber\\ && \times {\rm Tr} \left[i\Delta_{-+}(x,x') i\Delta_{-+}(x,x'''') iS_{++}(x',x'') iS_{++}(x'',x'''') iS_{++}(x'''',x''') iS_{++}(x''',x')i\Delta_{++}(x'',x''') \right. \nonumber\\ && \left.
+ i\Delta_{+-}(x,x') i\Delta_{+-}(x,x'''') iS_{--}(x',x'') iS_{--}(x'',x'''') iS_{--}(x'''',x''') iS_{--}(x''',x')i\Delta_{--}(x'',x''')\right] \nonumber\\ &&
= - \frac{3i g^4}{2^6 \pi^4}\int d^4 x' a'^2 \ln^2 a' \p'^{2}\left[  \left(i\Delta^2_{-+}(x,x')- i\Delta^2_{+-}(x,x')\right) \right] = \frac{3g^4H^2}{2^8 \pi^6}\ln^4 a.
\label{y36}
\end{eqnarray}

\bigskip 
\noindent
It is not difficult to see that the leading non-local contributions coming from both of \ref{f2} behaves as $\sim \ln^3 a $, which is subleading compared to the above.  The reason behind this can be understood as follows. The secular logarithms appearing in \ref{y28} originates from the renormalisation procedure, and both fermion and scalar propagators contribute into this. Once this has been evaluated, a massless fermion propagator, which is conformally flat, generates no further secular logarithms, whereas  the single scalar line generates one such logarithm.  Thus it is also clear that  $\ln^4 a$ appearing in \ref{y36} is basically a hybrid of local (coming from the self energy) and non-local (coming from the two external scalar lines) secular logarithms. \\

\begin{figure}[h!]
\begin{center}
  \includegraphics[width=11.5cm]{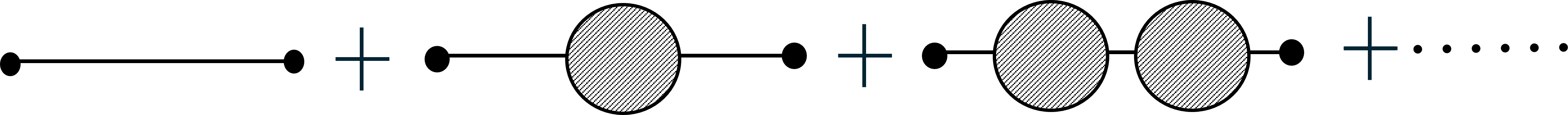}
 \caption{ \it \small Series summation  for the propagator due to infinite number of self energy loop insertions. For our present purpose, the two endpoints will be identified. The shaded circle represent the one plus two loop self energies containing leading secular terms (i.e. local in this case). }
  \label{seriesfig}
\end{center}
\end{figure}

\begin{figure}[h!]
\begin{center}
  \includegraphics[width=7.5cm]{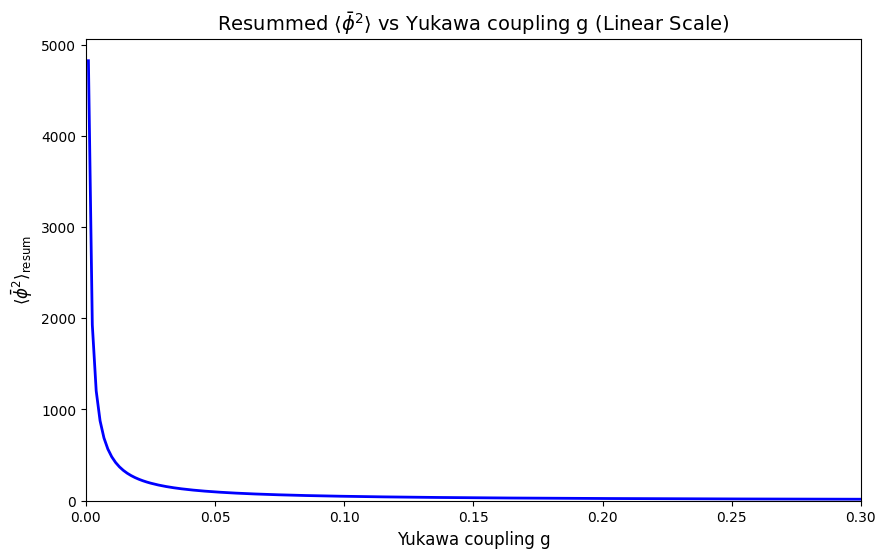}
 \caption{ \it \small The plot of the late time non-perturbative $\langle \bar{\phi}^2\rangle_{\rm resum}$ with respect to the Yukawa coupling. The same divergences for $g=0$, as is indicated by \ref{y37} or \ref{y38}. See main text for detail. }
  \label{plotfinal}
\end{center}
\end{figure}

\noindent
Putting things together now, the value of $\langle \phi^2\rangle$ reads in the leading order of the secular logarithm up to ${\cal O}(g^4)$,
\begin{eqnarray}
&& \langle \bar{\phi}^2\rangle = N - \frac{g^2 N^3}{3\pi^3} +\frac{3g^4 N^4}{64\pi^5},
\label{y37}
\end{eqnarray}
where we have abbreviated, $\bar{\phi} = 2\sqrt{\pi} \phi/H$ and $N= \ln a = Ht$. The above expectedly shows breakdown of perturbative expansion at late times. Hence let us try to make some resummation.  The standard chain resummation of \ref{y37} due to infinite series of self energy loop insertions  yields (\ref{seriesfig})
\begin{eqnarray}
&& \langle \bar{\phi}^2\rangle_{\rm resum} = \frac{N}{1 + \frac{g^2 N^2}{3\pi^3} -\frac{3g^4 N^3}{64\pi^5}}
\label{y38}
\end{eqnarray}
We may assign a non-perturbative value to $\langle \bar{\phi}^2\rangle_{\rm resum}$, by assigning the same to $N$ as follows.
We compute
\begin{eqnarray}
&& \frac{d\langle \bar{\phi}^2\rangle_{\rm resum}}{dN} = \frac{1- \frac{g^2 N^2}{3\pi^3} +\frac{3g^4 N^3}{32\pi^5} }{\left(1 + \frac{g^2 N^2}{3\pi^3} -\frac{3g^4 N^3}{64\pi^5}\right)^2}.
\label{y39}
\end{eqnarray}
  We next invert \ref{y38} in order to express $N$ as a function of $\langle \bar{\phi}^2\rangle_{\rm resum} $. Among the  three roots, the physical solution is the one which is positive, and satisfies  $N =0$ as $\langle \bar{\phi}^2\rangle_{\rm resum}=0$. We next write from \ref{y39},
\begin{eqnarray}
&& \int \frac{\left(1 + \frac{g^2 N^2(\langle \bar{\phi}^2\rangle_{\rm resum} )}{3\pi^3} -\frac{3g^4 N^3(\langle \bar{\phi}^2\rangle_{\rm resum})}{64\pi^5}\right)^2}{1- \frac{g^2 N^2(\langle \bar{\phi}^2\rangle_{\rm resum}) }{3\pi^3} +\frac{3g^4 N^3(\langle \bar{\phi}^2\rangle_{\rm resum} )}{32\pi^5} } d\langle \bar{\phi}^2\rangle_{\rm resum} =  N,
\label{y39}
\end{eqnarray}
The late time solution $(N \gg1)$, must correspond to the root of the denominator of the left hand side integration. This determines $\langle \bar{\phi}^2\rangle_{\rm resum}$. We must select the positive value of $\langle \bar{\phi}^2\rangle_{\rm resum}$ as the physical root. We have done this computation numerically and have plotted  $\langle \bar{\phi}^2\rangle_{\rm resum}$ in \ref{plotfinal}. The plot shows that the same decreases monotonically with the increasing Yukawa coupling. The above method is supposed to be an extension of the renormalisation group inspired autonomous methodology to resum de Sitter secular logarithms, proposed in~\cite{Kamenshchik:2020yyn, Kamenshchik:2021tjh, Kamenshchik:2024ybm}. Instead of inverting \ref{y37} maintaining perturbative order, we have worked with the resummed expression of \ref{y38}.
We can also determine the dynamically generated  mass of the scalar field through the non-perturbative value of $\langle \bar{\phi}^2\rangle_{\rm resum}$. It is well known that for a light scalar field, we have 
$$\langle {\phi}^2\rangle= \frac{3H^2}{8\pi m^2}$$
Substituting for the resummed value of $\langle \bar{\phi}^2\rangle_{\rm resum}$, and treating $m^2$ as $m^2_{\rm dyn.}$, we can compute the late time  dynamically generated mass of the scalar field due to the Yukawa interaction. Thus \ref{plotfinal} shows that the field becomes heavier with increasing value of the coupling parameter.  
\begin{figure}[h!]
\begin{center}
  \includegraphics[width=5.5cm]{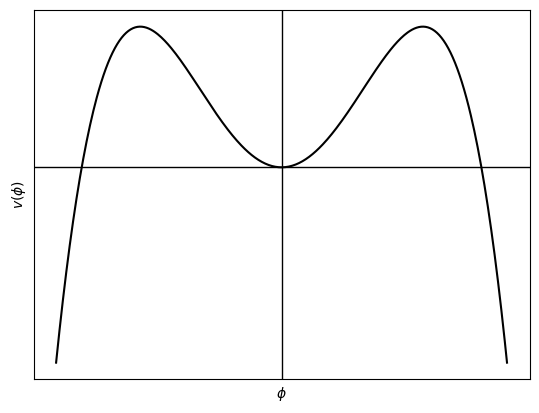}
 \caption{ \it \small Schematic diagram for the effective potential for the massless Yukawa theory in the absence of any scalar self interaction. For field vales away from zero, the potential becomes unbounded from below. See main text for discussion. }
  \label{effpot}
\end{center}
\end{figure}
Finally, we note that if one computes the effective potential for the Yukawa theory, the same becomes unbounded from below if there is no quartic self interaction with the lower bound  $\lambda \gtrsim 16 g^2$, e.g.~\cite{Bhattacharya:2025jtp} and references therein. Thus a finite value of $\langle \phi^2 \rangle$ and mass generation for such a theory might seem a bit  misleading. However, note that such unboundedness occurs away from $\phi =0$, as has been schematically depicted in \ref{effpot}. The analysis we have done in this paper takes the scalar to be purely quantum, i.e. no background field present,  and hence the scenario we are describing here is occurring around the minimum of  the potential. Nevertheless in the absence of any bounding effect, the field will eventually tunnel and escape the bounding region to enter an eternally rolling down phase.

\section{Discussion} \label{fin}
In this paper we have considered the two loop self energy for the scalar for the Yukawa interaction in the inflationary de Sitter spacetime. We have taken both the scalar and fermion to be massless. The one loop computations has been done in detail in earlier literatures. The chief motivation behind our present study corresponds to the fact that while the one loop scalar self energy contains only fermion lines, the two loop case, \ref{f2}, contain internal scalar lines. We have performed renormalisation in \ref{2loop}. We have next computed $\langle \phi^2 \rangle$ due to one and two loop self energies in \ref{NL-self}. We have argued that owing to the conformal invariance of massless fermion, the local self energy gives the leading secular behaviour compared to the non-local self energy. Next, a resummed expression for $\langle \phi^2\rangle$ has also been computed in the preceding section. The same decreases monotonically with increasing the magnitude of the Yukawa coupling.  Accordingly, the dynamically generated scalar mass increases with the increasing Yukawa coupling.

Perhaps a more realistic investigation in the context of primordial inflation will be to consider fermionic field theory in the presence of metric and the inflaton perturbation. This is supposed to automatically produce various interaction terms, often non-traditional, via the constraint equations (see e.g.~\cite{Seery}). Investigating loop effects generated via those interactions seems to be highly motivating as well as challenging. We hope to come to this issue sometime in the near future.

 \bigskip
\noindent
{\bf Acknowledgements :} The authors would like to acknowledge anonymous referee for some useful comments on an earlier version of this manuscript.


\bigskip
\appendix
\labelformat{section}{Appendix #1} 
\section{Some useful expressions}\label{A}
\noindent
Following~\cite{Brunier:2004sb}, we note for $d=4-\e$,
\begin{eqnarray}
\nonumber\\&&
\frac{1}{\Delta x^{6-4\e}}= - \frac{1}{2^2 \cdot 3  (2-2\e)(2-3\e)(1-2\e)\e} \p^4 \frac{1}{\Delta x^{2-4\e}}\nonumber\\&&
\frac{1}{\Delta x^{4-2\e}}= - \frac{1}{2 (1-\e)\e} \p^2 \frac{1}{\Delta x^{2-2\e}}
\label{A3}
\end{eqnarray}
We also note from \ref{props3'} that 
$$\p^2 \frac{1}{\Delta x_{++}^{2-\e}} =  \frac{4i \pi^{2-\e/2}}{\Gamma(1-\e/2)} \delta^d (x-x')$$
Thus from \ref{A3}, we have 
\begin{eqnarray}
\nonumber\\&&
\frac{1}{\Delta x^{4-2\e}_{++}}= -\frac{2i \mu^{-\e} \pi^{2-\e/2}}{(1-\e)\e \Gamma(1-\e/2)} \delta^d (x-x') -\frac{1}{2^2} \p^2 \frac{\ln \mu^2 \Delta x_{++}^2}{\Delta x_{++}^2} \nonumber\\&&
\frac{1}{\Delta x^{6-2\e}_{++}}= - \frac{i \mu^{-\e}\pi^{2-\e/2}}{(1-\e)(2-\e)^2\Gamma(1-\e/2)\e} \p^2 \delta^d(x-x')-\frac{1}{2^5} \p^4 \frac{\ln \mu^2 \Delta x_{++}^2}{\Delta x_{++}^2}
\label{A4}
\end{eqnarray}
where $\mu$ is an arbitrary  scale having the dimension of mass. Note in the above expressions, that the terms associated with the $\delta$-functions are local and carry divergences, whereas the logarithms are non-local. The $\delta$-functions arise from the $(|\eta-\eta'|\mp i\e)$ terms appearing in $\Delta^2_{++}(x,x')$ or $\Delta^2_{--}(x,x')$, \ref{props3'}. Thus   terms containing $\Delta x^2_{\pm \mp}$ cannot yield any local contribution or divergences, but they may contribute to the  infrared.

\bigskip

\end{document}